\documentclass{iaus}
\usepackage{graphicx}

\title[Magnetic Diffusion in Star Formation] 
{Magnetic Diffusion in Star Formation}

\author[Shantanu Basu \& Wolf B. Dapp]   
{Shantanu Basu \and Wolf B. Dapp} 

\affiliation{Department of Physics and Astronomy, The University of Western Ontario, \\ 
London, Ontario N6A 3K7, Canada\\ email: {\tt basu@uwo.ca; wdapp@uwo.ca}}

\pubyear{2010}
\volume{270}  
\pagerange{XXX--XXX}
\jname{Computational Star Formation}
\editors{J. Alves, B. G. Elmegreen, J. M. Girart, \& V. Trimble, eds.}
\begin{document}

\maketitle

\begin{abstract}
Magnetic diffusion plays a vital role in star formation.
We trace its influence from interstellar cloud scales
down to star-disk scales. 
On both scales, we find that magnetic diffusion can be significantly 
enhanced by the buildup of strong gradients in magnetic field structure.
Large scale nonlinear flows can create compressed cloud layers within which 
ambipolar diffusion occurs rapidly. However, in the flux-freezing limit
that may be applicable to photoionized molecular cloud envelopes, supersonic
motions can persist for long times if driven by an externally generated magnetic field
that corresponds to a subcritical mass-to-flux ratio.
In the case of protostellar accretion, rapid magnetic diffusion (through
Ohmic dissipation with additional support from ambipolar diffusion) near the protostar
causes dramatic magnetic flux loss. By doing so, it also allows the formation of
a centrifugal disk, thereby avoiding the magnetic braking catastrophe.
 
\end{abstract}

\firstsection 
\section{Clouds to Cores}

\begin{figure}[b]
\begin{center}
 \includegraphics[width=1.7in]{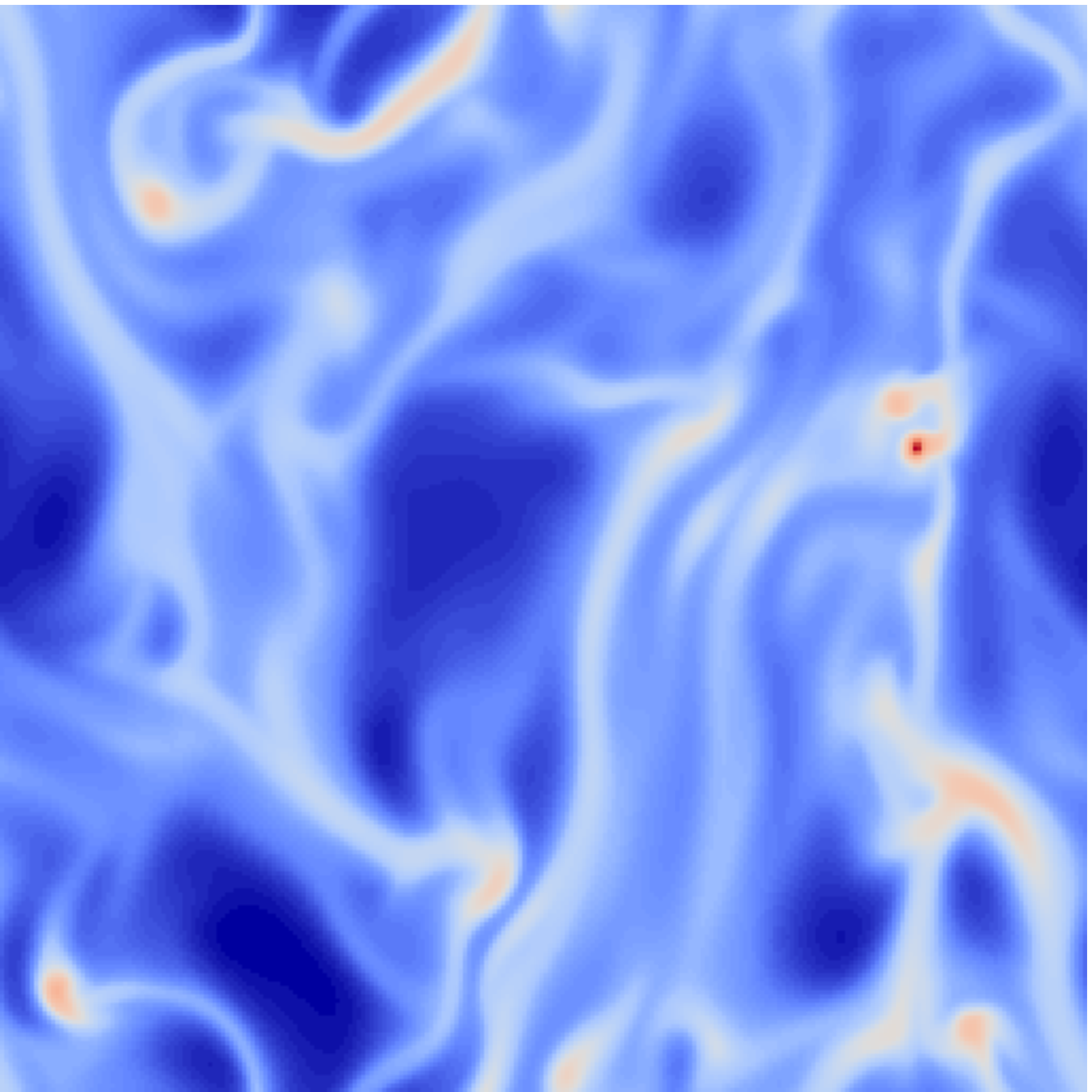} 
 \includegraphics[width=1.7in]{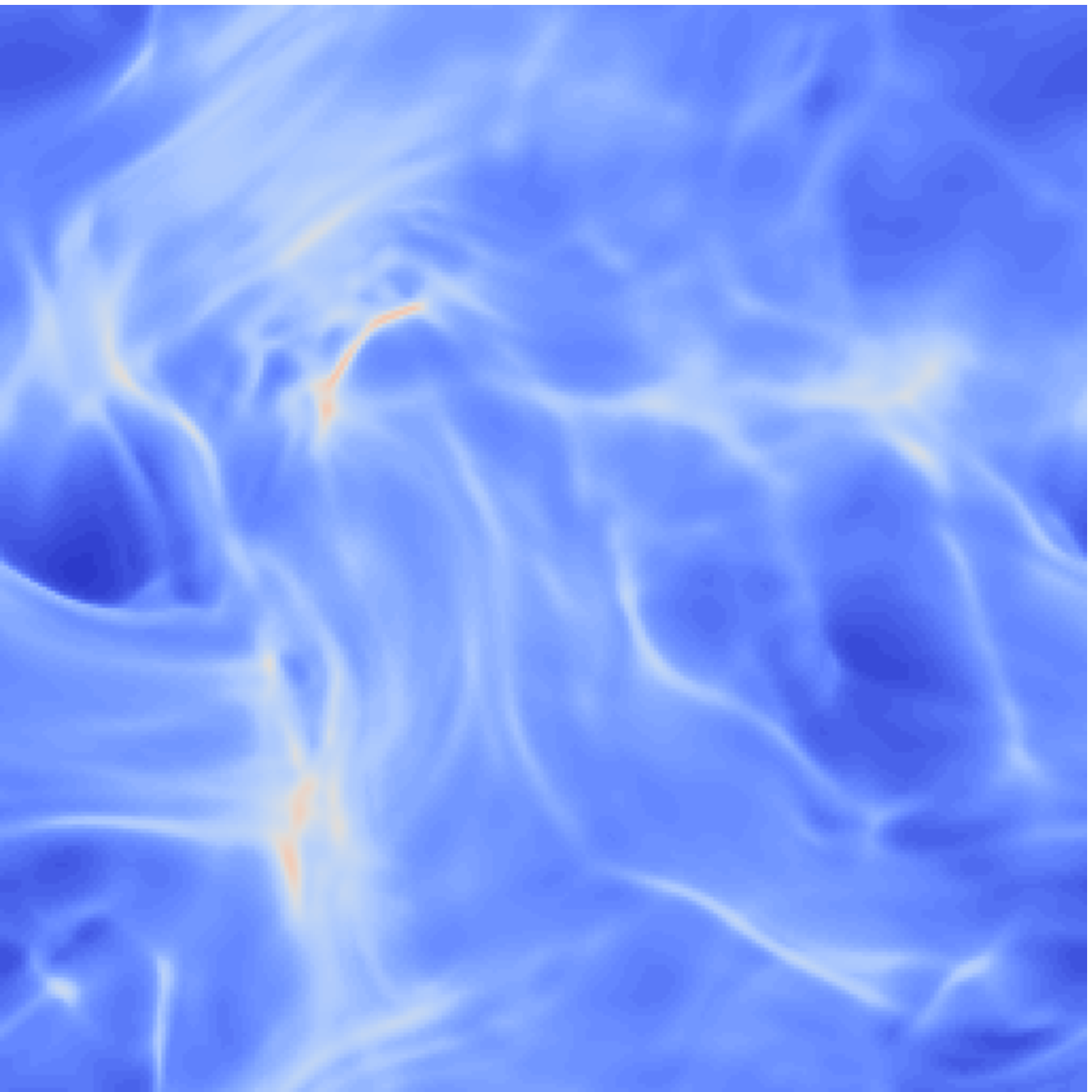} 
 \caption{Images of gas column density for initially turbulent models with ambipolar diffusion (left) and 
flux freezing (right), shown in identical logarithmic color schemes.
Both models have initially the same subcritical mass-to-flux ratio and turbulent power spectrum and amplitude,
that is allowed to decay. Both models are shown at the same physical time, with the model on the left undergoing runaway 
collapse in isolated cores and the model on the right in the midst of indefinite supersonic motions. 
From \cite[Basu \& Dapp (2010)]{bas10}.
}
   \label{fig1}
\end{center}
\end{figure}

\begin{figure}[b]
\begin{center}
 \includegraphics[width=2.5in]{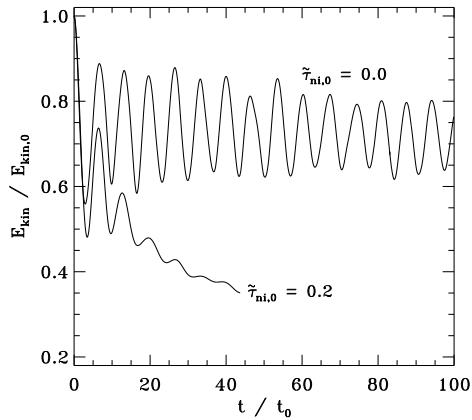} 
 \caption{Decay of kinetic energy. The time evolution of total kinetic energy,
normalized to its initial value, for a model with flux freezing ($\tilde{\tau}_{\rm ni,0}=0$)
and another with ambipolar diffusion corresponding to a canonical ionization fraction
for molecular clouds ($\tilde{\tau}_{\rm ni,0}=0.2$). 
From \cite[Basu \& Dapp (2010)]{bas10}. }
   \label{fig2}
\end{center}
\end{figure}

Magnetic energy dominates self-gravitational energy in the H {\small I} clouds that
occupy the interstellar medium (ISM) of our Galaxy 
(\cite[Heiles \& Troland 2005]{hei05}). In other words, their
mass-to-flux ratio $M/\Phi$ is significantly less than the critical value
required for gravitational collapse and fragmentation.
On the other hand, molecular clouds, the birthplaces of stars, have 
mass-to-flux ratios that are very close to the critical value
(\cite[Crutcher 2004]{cru04}).
Mass-to-flux ratios in molecular clouds (or cloud fragments)
that are significantly greater than the ambient ISM value can be achieved by
two distinct but not mutually exclusive mechanisms. One is the accumulation of matter 
{\it along} the ambient magnetic field direction in the local spiral arm.
However, this places severe constraints on the accumulation length of 
molecular clouds, and on either the associated formation timescale or the magnitude
of streaming motions that can create the clouds (\cite[Mestel 1999]{mes99}).   
The second and rather attractive possibility is that the formation process of the
molecular cloud, in a medium pervaded by turbulence or nonlinear
flows, will lead to rapid ambipolar diffusion in at least the compressed
(filamentary or sheet-like) regions. This scenario of turbulent ambipolar diffusion
has been explored in several recent works, e.g., \cite[Li \& Nakamura (2004)]{li04}, \cite[Kudoh \& Basu (2008)]{kud08}, and \cite[Basu et al. (2009)]{bas09}.

\begin{figure}[b]
\begin{center}
 \includegraphics[width=2.5in]{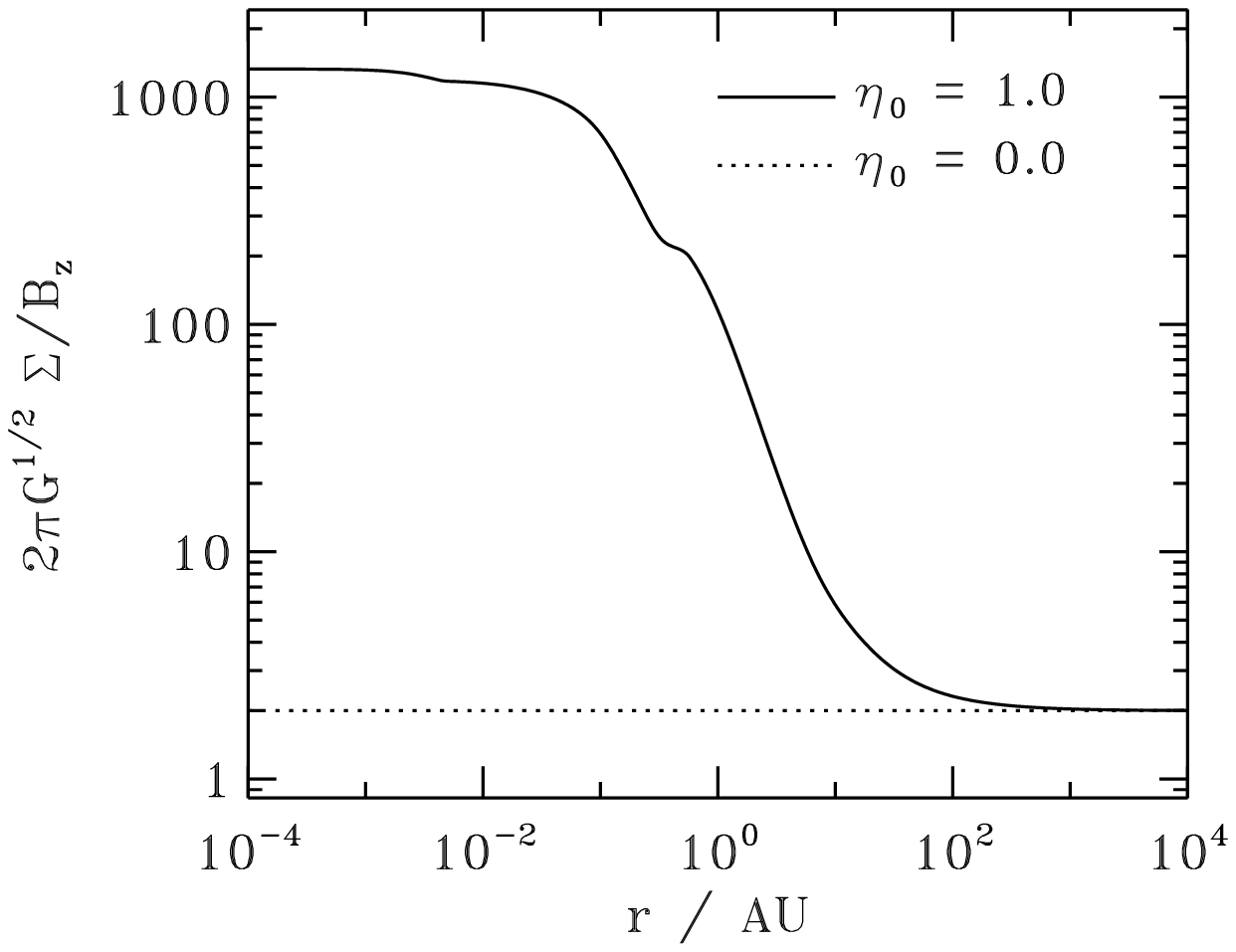} 
 \includegraphics[width=2.5in]{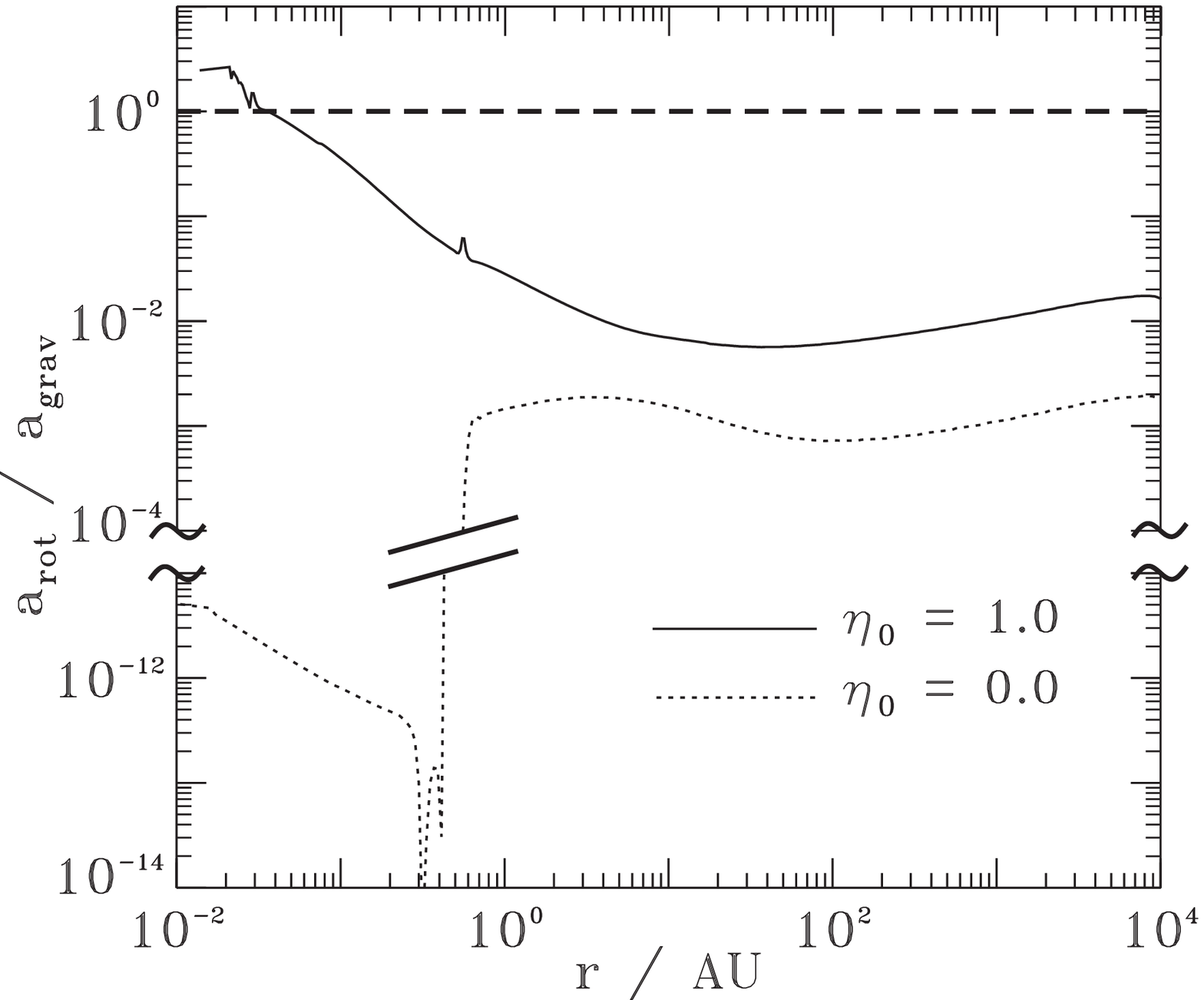} 
 \caption{Spatial profiles of mass-to-flux ratio (left) and centrifugal support level (right).}
Both profiles are shown after the formation of the second core, with extent 
$\sim R_{\odot}$. The plot on the right is shown at a time shortly 
after the introduction of a central sink cell of radius $3\, R_{\odot}$
that masks the newly-formed second core.
From \cite[Dapp \& Basu (2010)]{dap10}.
   \label{fig3}
\end{center}
\end{figure}

\begin{figure}[b]
\begin{center}
 \includegraphics[width=1.7in]{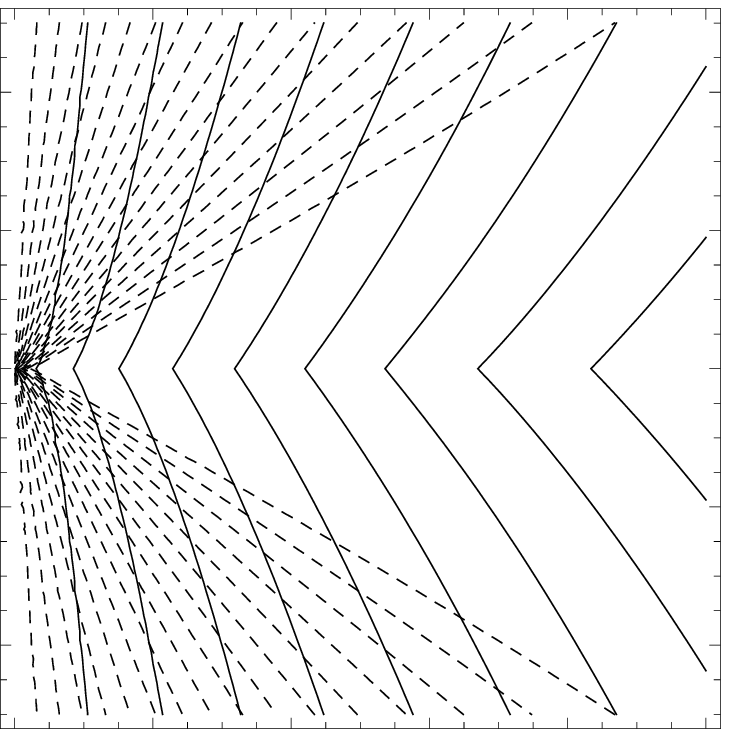} 
 \includegraphics[width=1.7in]{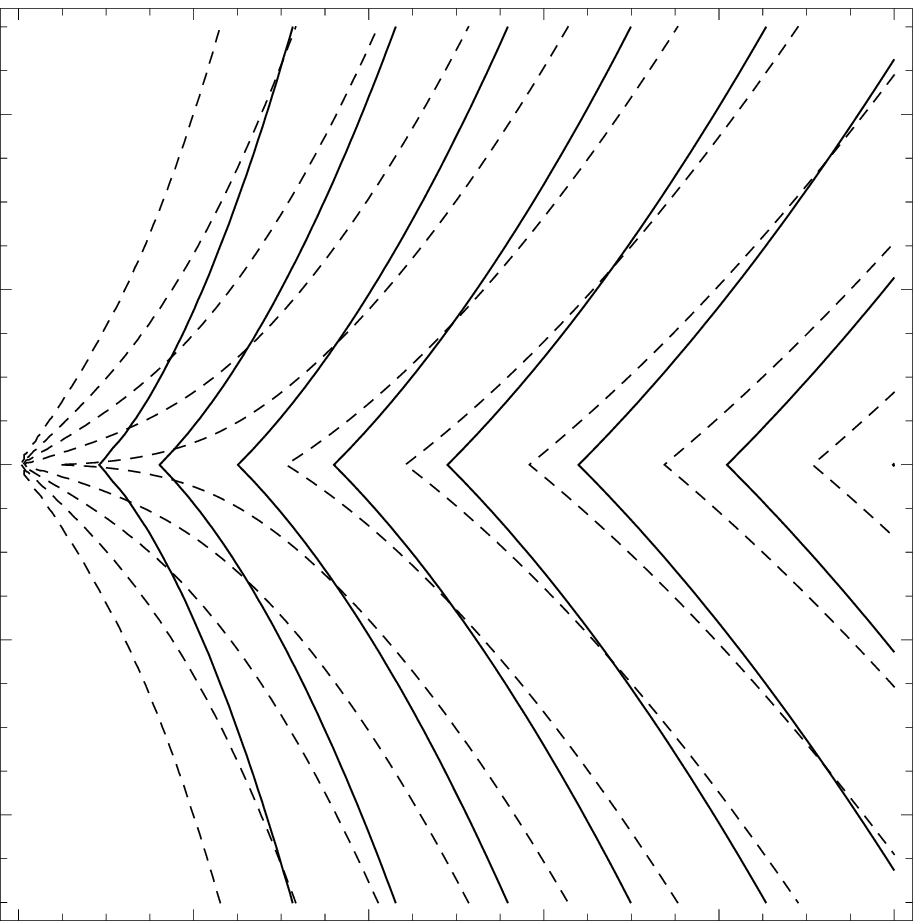} 
 \caption{Magnetic field lines. The box on the left has dimensions
10 AU on each side, while the box on the right has dimensions 100 AU on
each side. The dashed lines represent the flux-freezing model
($\eta_0=0$) while the solid lines show the same field lines for
$\eta_0=1$. The second core has just formed and is on the left axis midplane.
}
   \label{fig4}
\end{center}
\end{figure}

Here, we report the results of \cite[Basu \& Dapp (2010)]{bas10}. 
Solution of the thin-sheet MHD equations, including ambipolar diffusion,
shows that the rapid accumulation of subcritical gas is expected to lead to 
islands of higher mass-to-flux ratio. These regions undergo enhanced ambipolar 
diffusion during the compression, and if they are still 
subcritical, they then undergo 
ambipolar-diffusion-driven contraction more rapidly than their 
surroundings due to the elevated density 
in those regions. The qualitative result is a handful of cores that are formed within 
elongated ridges.
These cores undergo runaway collapse with subsequent near-flux-trapping as
soon as gravity overwhelms magnetic and thermal pressure forces within a
supercritical region.
However, most of the gas in the cloud remains subcritical, does not form
stars, and maintains a higher velocity dispersion.
This mode of star formation is illustrated in the left panel of Fig. 1.
The dark (red) dots represent high density regions of cores that are 
undergoing runaway collapse. This occurs at a time $43.5\, t_0$ in this model, 
where $t_0 = c_{\rm s}/(2 \pi G \Sigma_0)$, $c_{\rm s}$ is the
isothermal sound speed, and $\Sigma_0$ is the mean column density of the sheet.
The collapse time is a factor $\approx 6$ shorter than if starting with small-amplitude
initial perturbations.
The right panel shows the state of evolution at the same physical time
and for statistically identical initial conditions but with ambipolar diffusion turned off,
i.e., pure flux freezing. In this case, a startling result is that 
supersonic motions persist indefinitely. This follows an initial phase
in which some but not all of the initial kinetic energy is lost due to 
shocks. A full explanation and theory
is given by \cite[Basu \& Dapp (2010)]{bas10}.
Fully three-dimensional simulations by \cite[Kudoh \& Basu (2008)]{kud08}
confirm the rapid core formation found in thin-sheet models such as this and
those of \cite[Li \& Nakamura (2004)]{li04} and \cite[Basu et al. (2009)]{bas09}.
The thin-sheet flux-freezing result of long-lived supersonic motions 
(\cite[Basu \& Dapp 2010]{bas10}) remains to be
tested fully in three dimensions since the result is dependent upon
the existence of a low density external medium in which the magnetic field
can quickly adjust to a current-free configuration. The latter is built in to
the thin-sheet models and may be approximately true in a three-dimensional
model of a cold magnetized cloud embedded in a hot tenuous medium.

\section{Cores to Star-Disk Systems}

Core collapse inevitably begins when the mass-to-flux ratio is 
a factor $\approx 2$ above the critical value (\cite[e.g., Basu \&
Mouschovias 1994]{bas94}).
Rapid collapse on a dynamical timescale is able to effectively
trap the remaining magnetic flux during the prestellar runaway 
collapse phase. If this flux trapping continued indefinitely, there
would remain a big magnetic flux problem for the final star.
A cloud core with twice the critical mass-to-flux ratio
would still contain $10^8$ times as much magnetic flux per mass as threads
the solar surface, and $10^3-10^5$ times as much as a 
magnetic Ap star or a T Tauri star. 

Fortunately, a resolution to the magnetic flux problem is facilitated
in the post-stellar-core formation epoch, also known as the
accretion phase or ``$t>0$", where $t=0$ is the pivotal moment at
which a central protostar is formed. In the spherically 
symmetric model of \cite[Shu (1977)]{shu77}, $t=0$ is pivotal in that an 
expansion wave subsequently moves outward from the center and envelopes 
a region of near-free-fall that is dominated by the potential of the 
central point mass.
For the purpose of the magnetic flux problem, $t=0$ is also pivotal.
The subsequent collapse leads to sharp magnetic field gradients
in the innermost regions 
such that the diffusion terms dominate the advection term in the  
magnetic induction equation, as shown by 
\cite[Li \& McKee (1996)]{li96} and 
\cite[Contopoulos et al. (1998)]{con98}. Expressed more loosely, the 
extreme dragging-in of magnetic field lines in the flux-freezing 
limit leads to a split-monopole configuration that would in reality
be prone to rapid magnetic diffusion due its sharp magnetic field gradient.
It is essentially a self-regulation by magnetic diffusion that prevents a 
split-monopole from forming in the accretion phase.

Another classical problem of star formation is due to angular momentum.
Even a rotation rate of $\sim 10^{-14}$ rad s$^{-1}$ as observed in molecular
clouds (\cite[Goodman et al. 1993]{goo93}), 
while not dynamically important on cloud scales, contains enough
angular momentum to prevent nearly all the matter from falling in to 
a central region of stellar dimensions.
However, collapse in the prestellar phase is never sufficient to
raise the level of centrifugal support relative to gravity, a 
result decisively shown by the simulations of 
\cite[Norman et al. (1980)]{nor80}
and explained analytically as a property of self-similar collapse 
profiles by \cite[Basu \& Mouschovias (1995)]{bas95}.
Magnetic braking further weakens the level of centrifugal support, 
primarily in the core formation epoch, because the subsequent runaway collapse of a
prestellar core is generally too rapid for magnetic braking to remain
active (\cite[Basu \& Mouschovias 1994]{bas94}). 
Here again, $t=0$ provides a pivot point, after which a centrifugal 
barrier {\it does} exist for infalling mass shells, as they fall inward
under the gravitational domination of a central protostar.
While this implies that a disk will be formed (only) in the $t>0$ phase,
\cite[Allen et al. (2003)]{all03} in fact found that magnetic braking 
gets revitalized in this phase under the assumption of flux freezing. 
For $t>0$, the extreme flaring of the magnetic field due to a monopole-like
configuration is able to couple near-stellar regions to regions
of far greater moment-of-inertia, leading to very efficient magnetic
braking. Also, even if a centrifugal disk begins to form during the
$t>0$ phase, the radial velocity is slowed down enough that magnetic braking 
has time to act and prevent the ultimate formation of the disk.

We have solved the MHD equations for a rotating thin sheet, 
including the effect of Ohmic 
dissipation, in axisymmetric geometry. Our results published in 
\cite[Dapp \& Basu (2010)]{dap10} show that significant magnetic flux loss occurs 
within the first core (radius $\sim$ few AU), and effectively shuts down magnetic braking. 
Ohmic dissipation is added according to the prescription of
\cite[Nakano et al. (2002)]{nak02}, similar to the implementation of
\cite[Machida et al. (2007)]{mac07}, but accounting for spatial gradients
in the resistivity. A dimensionless scaling parameter $\eta_0$
characterizes the resistivity, in which the uncertainties hinge largely on the 
grain size distribution. A canonical value is $\eta_0=1$. 
Additional significant magnetic diffusion can occur at high densities due to 
ambipolar diffusion, as studied by \cite[Kunz \& Mouschovias (2010)]{kun10},
but is not included in our present study.
Ohmic dissipation and its facilitation of disk formation has also been modeled
by \cite[Krasnopolsky et al. (2010)]{kra10} but only on larger scales 
( $> 10$ AU).

The eventual transformation
of a first core to a second core of radius $\sim\,R_{\odot}$, due to H$_2$ dissociation,
 then occurs with near angular
momentum conservation. This is due to rapid Ohmic dissipation having rendered 
magnetic braking ineffective in this region. The result is that a 
centrifugal disk does indeed form in the near-environment of the newly-formed 
second core with mass only $\sim\,10^{-3}\, M_{\odot}$. 
The rapid flux loss (as seen by a dramatically increased mass-to-flux
ratio) is shown in the left panel of Fig. 3. The right panel illustrates the
achievement of centrifugal
balance in the late stages of the resistive model ($\eta_0=1$), and
the {\it magnetic braking catastrophe} in the flux-freezing model ($\eta_0=0$).
In the latter case, centrifugal support is decimated 
in the innermost regions by magnetic braking, and a disk cannot form.

Fig. 4 shows the dramatic difference in magnetic field line structure above and 
below the sheet, calculated using the current-free approximation from a scalar
magnetic potential. The split monopole of the $\eta_0=0$ model is replaced by 
a much more relaxed field line structure. 
The extreme flaring of field lines in the $\eta_0=0$ model is a fundamental cause of
the magnetic braking catastrophe. More details are in 
\cite[Dapp \& Basu (2010)]{dap10}.
Similar results on magnetic field structure, using a simplified model of
Ohmic dissipation, can be found in \cite[Galli et al. (2009)]{gal09}.

\end{document}